\documentclass[11pt]{article}
\newcommand{\be}{\begin{equation}}
\newcommand{\ee}{\end{equation}}
\newcommand{\bea}{\begin{eqnarray}}
\newcommand{\eea}{\end{eqnarray}}

\newcommand{\sn}{{\rm sn}}
\newcommand{\cn}{{\rm cn}}
\newcommand{\dn}{{\rm dn}}

\newcommand{\m}{{\rm m}\,}

\newcommand{\sss}{{\vspace{.2in}}}

\begin{document}
~\hfill{\footnotesize SUNYB/04-02, IOP-BBSR/04-04}
\vspace{.5in}
\begin{center}
{\LARGE {\bf Analytically Solvable PT-Invariant Periodic Potentials}}
\end{center}
\vspace{.3in}
\begin{center}
{\Large{\bf  \mbox{Avinash Khare}
 }}\\
\noindent
{\large Institute of Physics, Sachivalaya Marg, Bhubaneswar 751005, India}
\end{center}
\begin{center}
{\Large{\bf  \mbox{Uday Sukhatme}
 }}\\
\noindent
{\large Department of Physics, State University of New York at Buffalo, Buffalo, NY 14260, U.S.A. }\\
\end{center}
\vspace{.9in}
{\bf {Abstract:}}  
Associated Lam\'e potentials $V(x)=a(a+1)m\sn^2(x,m)+b(b+1)m{\cn^2 (x,m)}/{\dn^2(x,m)}$ are used 
to construct complex, PT-invariant, periodic potentials using the anti-isospectral
transformation $x \rightarrow ix+\beta$, where $\beta$ is any nonzero real number. 
These PT-invariant potentials are defined by $V^{PT}(x) \equiv -V(ix+\beta)$, and have a different 
real period from $V(x)$. They are analytically solvable potentials with a finite number of band gaps, 
when $a$ and $b$ are integers.  
Explicit expressions for the band edges of some of these potentials
are given. For the special case of the complex potential
$V^{PT}(x)=-2m\sn^2(ix+\beta,m)$, we also analytically obtain the
dispersion relation. Additional new, solvable, complex, PT-invariant, periodic potentials 
are obtained by applying the techniques of supersymmetric quantum mechanics. 

\newpage

In the past few years, Bender and others \cite{bed,oth} 
have looked at several complex potentials
with PT-symmetry and have shown that the energy eigenvalues 
are real when PT-symmetry is unbroken, whereas they come in complex
conjugate pairs when PT-symmetry is spontaneously broken. 
Recently, Mostafazadeh \cite{mos} has clarified this issue by showing 
that there is a more general setting of pseudo-hermiticity (of which
PT-symmetry forms a special case) in which eigenvalues are either real 
or occur in complex conjugate pairs. Scattering problems with complex, PT-invariant potentials
have also been investigated \cite{bdr,ah1}. However, there have been very few papers discussing
periodic potentials with PT-symmetry. Only two types of PT-invariant, periodic potentials 
have been considered in detail in the
literature, namely, $i\sin^{2N+1}(x)$ \cite{bdm} and delta function
potentials with complex couplings  \cite{za,cer}. These potentials have been shown to possess 
real band spectra with an infinite number of band gaps. It might
be noted here that obtaining these results required extensive numerical analysis and 
no analytic results for band edge energies were possible \cite{km}. 
The purpose of this letter is to construct and study several new classes of analytically
solvable, complex, PT-invariant, periodic potentials with a finite number of band
gaps. Our approach will consist of (i) making use of the anti-isospectral  
transformation $x \rightarrow ix+\beta$  \cite{kra}, and (ii) constructing supersymmetric 
partner potentials from techniques developed in supersymmetric quantum mechanics \cite{cks}.

\noindent {\bf Anti-Isospectral Transformations:} We begin with the simple observation 
that if $\psi(x)$ is a solution of the Schr\"odinger equation for the real potential 
$V(x)$ with energy $E$, then $\psi(ix+\beta)$ is a solution of the Schr\"odinger equation 
for the complex potential $-V(ix+\beta)$ with energy $-E$, where $\beta$ is an arbitrary 
constant. The new potential $-V(ix+\beta)$, generated by the anti-isospectral 
transformation $x \rightarrow ix+\beta$, is clearly PT-symmetric and will be denoted 
by $V^{PT}(x)$. Further, if $\psi(x)$ and $\psi(ix+\beta)$ satisfy appropriate boundary 
conditions, they are eigenfunctions of $V(x)$ and $V^{PT}(x)$ respectively. Since the 
ordering of energy levels for $V^{PT}(x)$ is the opposite of the ordering of 
energy levels for $
V(x)$, this is presumably why the transformation $x \rightarrow ix+\beta$ is 
called ``anti-isospectral".

In this letter, we focus on periodic potentials by choosing $V(x)$ 
to be the associated Lam\'e potential
\be\label{eq1}
V(x)=a(a+1)m\sn^2(x,m)+b(b+1)m\frac{\cn^2 (x,m)}{\dn^2(x,m)}\,,
\ee 
which has a real period $2K(m)$. Note that if either $a$ or $b$ is zero, this 
potential is called the Lam\'e potential. Recall that when $a$ and $b$ are non-negative 
integers, the associated Lam\'e potential has many analytically solvable eigenstates and 
only a finite number of band gaps \cite{ks1,ks2}. 
Here, $\sn \,(x,\m)$, $\cn \,(x,\m)$, $\dn \,(x,\m)$ are Jacobi elliptic 
functions with elliptic
modulus parameter $\m$ $( 0\leq \m\leq 1)$. They are doubly periodic
functions with periods $[4K(\m), i2K'(\m)]$, $[4K(\m), 2K(\m)+i2K'(\m)]$,
$[2K(\m), i4K'(\m)]$ respectively \cite{abr,gr}, where 
$K(\m) \equiv \int_0^{\pi/2} d\theta [1-\m\sin^2 \theta]^{-1/2}$ 
denotes the complete elliptic integral of the first kind, and
$K'(\m)\equiv K(1-\m)$. The complex, PT-invariant potential obtained via an
anti-isospectral transformation applied to eq. (\ref{eq1}) is
\be\label{eq2}
V^{PT}(x)=-a(a+1)m\sn^2(ix+\beta,m)-b(b+1)m\frac{\cn^2 (ix+\beta,m)}{\dn^2(ix+\beta,m)}\,.
\ee   
It is also periodic with a different real period $2K'(m)$. Furthermore, it is  
analytically solvable with a finite number of band gaps. 
It is important to understand that the key point for obtaining the above results
is that unlike
trigonometric and other periodic functions, Jacobi elliptic
functions are doubly periodic functions. This allows both $V(x)$ and $V^{PT}(x)$ 
to be simultaneously periodic, even though the periods are different. Note that the arbitrary 
nonzero constant $\beta$ in the anti-isospectral transformation,  $x \rightarrow ix+\beta$, 
is chosen so as to avoid the singularities of Jacobi elliptic functions \cite{abr}. 

Let us first apply our approach to the Lam\'e potentials ($b=0$)
\be\label{eq3}
V(x)=a(a+1)m\sn^2(x,m)\,,~~a=1,2,3,...\,, 
\ee
which are known to have $2a+1$ eigenstates (band edges) and $a$ band gaps. Let $E_j(m)$ 
and $\psi_j(x,m)$ with $j=0,1,\ldots,2a$ denote the band edge energies and wave functions.     
The anti-isospectral transformation $x \rightarrow ix+\beta$ 
\cite{kra} yields the
PT-invariant potential 
\be\label{eq4}
V^{PT}(x)=-a(a+1)m\sn^2(ix+\beta,m)\,,~~a=1,2,3,...\,, 
\ee
with real period $2K'(m)$. The band-edge eigenvalues and eigenfunctions of $V^{PT}(x)$ 
are related to those of the Lam\'e potential (\ref{eq3}) by 
\be\label{eq5}
E_j^{PT}(m) =
-E_{2a-j}(m)\,,~~~\psi_j^{PT}(x,m)~\propto~\psi_{2a-j}(ix+\beta,m)~,~~~j=0,1,...
,2a\,.
\ee  
Thus, the PT-invariant, periodic potential (\ref{eq4}) also has
precisely $a$ band gaps and $2a+1$ band edges at energies 
given by eq. (\ref{eq5}). Special mention should be made of the
remarkable fact that for any integer $a$, all  bands and band gaps 
exchange their role as one goes from a Lam\'e potential 
to its PT-invariant version $V^{PT}(x)$.

For any band structure problem, an important quantity is the
discriminant $\Delta$ \cite{mw} which gives information
about the number of band gaps as well as their widths. The question is whether one can 
relate the discriminant  for the potential $V^{PT}(x)$ with
the discriminant $\Delta$ 
for the corresponding Lam\'e potential. Unfortunately, this is not directly possible by 
using eq. (\ref{eq5}), since 
it only relates energies of states with different numbers of
nodes. However, we now derive a remarkable relation using which we can relate
the two discriminants $\Delta$ and $\Delta^{PT}$.

We start from the Schr\"odinger equation for the Lam\'e potential
(\ref{eq3})
\be\label{eq6}
-\psi''(x)+a(a+1)m\sn^2(x,m) \psi(x) = E(m) \psi(x)\,,
\ee
where a prime denotes a derivative with respect to the argument. On using
the relation \cite{abr,gr}
\be\label{eq7}
\sqrt{m}\sn(x,m)=-\dn[ix+K'(m)+iK(m),1-m]\,,
\ee
and then defining a new variable $y=ix+K'(m)+iK(m)$, the
Schr\"odinger eq. (\ref{eq6}) takes the form 
\be\label{eq8}
-\psi''(y)+a(a+1)(1-m)\sn^2(y,1-m) \psi(y) = [a(a+1)-E(m)] \psi(y)\,,
\ee
so that for the Lam\'e potentials (\ref{eq3}) we obtain the remarkable
relations
\be\label{eq9}
E_j (m) = a(a+1) -E_{2a-j} (1-m)\,,~\psi_j(x,m)\propto\psi_{2a-j}(ix+K'(m)+iK(m), 1-m),~~j=0,1,...,2a.
\ee
In passing, note that for the special choice $m=1/2$, one has several interesting relations:  
\be\label{eq10}
E_j (m=1/2) +E_{2a-j} (m=1/2) = a(a+1)\,,~~ E_{a} (m=1/2) =
a(a+1)/2\,.
\ee
On combining eqs. (\ref{eq5}) and (\ref{eq9}) we obtain
\be\label{eq11}
E^{PT}_j (m) = E_{j} (1-m) -a(a+1)\,,~~j=0,1,...,2a\,,
\ee
and hence the corresponding discriminants are  
related by 
\be\label{eq12}
\Delta^{PT} (E,m)=\Delta[E+a(a+1),1-m]\,. 
\ee

As an illustration, in Figure 1 we plot the real and
imaginary parts of the PT-invariant, complex potential 
$V^{PT}(x)=-12m\sn^2(ix+\beta,m)$. Using the well known results for the Lam\'e 
potential with $a=3$ \cite{ks1} and eq. (\ref{eq5}), 
the ground state (lowest band edge)
eigenvalue and eigenfunction is
\be\label{eq13}
\psi_g(x)=\sn(ix+\beta,m)[2+2m-\delta_3-5m\sn^2(ix+\beta,m)]\,~,~  
E_g = -5-5m-2\delta_3\,,
\ee
where $\delta_3 \equiv \sqrt{4-7m+4m^2}$.
In Table I we have given all the seven  
band edge eigenvalues and eigenfunctions. We have subtracted off
the ground state energy from the potential so that the lowest band
edge by construction is at zero energy. Observe from the table that
the band edges are both periodic as well as anti-periodic with periods
$2K'(m)$ and $4K'(m)$ respectively.  

For the special case of the Lam\'e potential with $a=1$, the dispersion relation is also
analytically known \cite{ww}. We
now obtain the dispersion relation for the corresponding  
PT-invariant potential $V^{PT}(x)=-2m\sn^2(ix+\beta,m)$. To that
end, we start from the Schr\"odinger equation: 
\be\label{eq14}
-\psi''(x)+[1+m-2m\sn^2(ix+\beta,m)] \psi(x) = E \psi(x)\,,
\ee
where we have subtracted the ground state energy $E_g = -1-m$ from the potential so
that the new potential $[V^{PT}]_-(x) = V^{PT}(x)-E_g$ has zero ground state energy. On substituting
$y=ix+\beta$, eq. (\ref{eq14}) takes the form
\be\label{eq15}
-\psi''(y)+[-m+2m\sn^2(y,m)] \psi(y) = (1-E) \psi(y)\,.
\ee
Now it is well known that two independent solutions of this equation
are given by \cite{ww}
\be\label{eq16}
\psi(x)=\frac{H(ix+\beta\pm \alpha_1)\exp[\mp (ix+\beta) Z(\alpha_1)]}
{\theta(ix+\beta)}\,,
\ee
where $H,\theta,Z$ are the Jacobi eta, theta and zeta functions, while
$\alpha_1$ is related to the energy $E$ of eq. (\ref{eq15}) by 
\be\label{eq17}
E=m\sn^2(\alpha_1,m)\,.
\ee
On using the Bloch condition and the fact that while 
$\theta(ix+\beta)$ is a periodic function with period 
$2K'(m)$, $H(ix+\beta)$ is only quasi-periodic \cite{ww}, i.e.
\be\label{eq18}
H(i[x+2K'(m)]+\beta)=H(ix+\beta)\exp[-\pi K'(m) /K(m)]\,,
\ee
it is easily shown that the
complex, PT-invariant potential $[V^{PT}]_-(x)$ has a dispersion relation
given by
\be\label{eq19}
k = \mp \frac{\pi}{2K'(m)} \pm iZ(\alpha_1) + i\frac{\pi}{2K(m)}\,,
\ee
where $\alpha_1$ is given by eq. (\ref{eq17}). 

We now turn to the associated Lam\'e potentials of eq.
(\ref{eq1}), where without loss of generality we consider $a > b$ with
both being positive integers. We shall later comment about the case $a=b$. 
As has been shown by us \cite{ks1,ks2}, these are also exactly 
solvable problems with precisely $a$ band
gaps and $2a+1$ band edges. However, in many of these cases, some of the
bands are unusual in that both the band edges have the
same period, since some band gaps vanish. In this sense, the associated Lam\'e 
potentials are much richer than the Lam\'e potentials.  
On using the anti-isospectral transformation, it is easy to
see that the band edges of the potential (\ref{eq1}) and
its PT-invariant counterpart  
\be\label{eq20}
V^{PT}(x)=-a(a+1)m\sn^2(ix+\beta,m)-b(b+1)m\frac{\cn^2 (ix+\beta,m)}
{\dn^2(ix+\beta,m)}\,,
\ee
are again connected by the relation (\ref{eq5}). However, we are unable to
relate the discriminants of the two potentials since we have not
been able to derive an  analogue of
the relation (\ref{eq9}).  As an illustration, let us consider the ($a$=2, $b$=1)
associated Lam\'e potential and its corresponding PT-invariant potential 
$V^{PT}(x)=-6m\sn^2(ix+\beta,m)-2m{\cn^2 (ix+\beta,m)}/{\dn^2(ix+\beta,m)}\,$. 
The ground state eigenvalue and eigenfunction is given by
\be\label{eq21}
\psi_g (x)= \frac{\cn(ix+\beta,m}{\dn(ix+\beta,m)}
[3m\sn^2(ix+\beta,m)-2+\sqrt{4-3m}]\,,~~E_g = -5-m-2\sqrt{4-3m}\,.
\ee
In Table 2 we have given all the band edge eigenvalues
and eigenfunctions.

Finally, let us discuss the associated Lam\'e potentials (\ref{eq1}) for the case
$a=b$ = integer.  In view of the well known Landen transformation formula \cite{abr,ks3}
\be\label{eq22}
\dn(x,m)+\dn(x+K(m),m)=\frac{1}{\alpha}\dn \left[\frac{x}{\alpha},\tilde{m} \right]\,,
~~\alpha = \frac{1}{1+\sqrt{1-m}}\,,~~\tilde{m} = \left[\frac{1-\sqrt{1-m}}
{1+\sqrt{1-m}}\right]^2\,,
\ee
the associated Lam\'e potentials with $a=b$ can be rewritten,
apart from an overall constant as 
Lam\'e potentials and so the results derived above for the Lam\'e
potentials will go through with a modified modulus parameter $\tilde{m}$.

\noindent {\bf Supersymmetric Partner Potentials:} Additional analytically 
solvable finite band gap potentials can be
obtained from our previous results by using supersymmetry. 
The procedure is standard \cite{cks}. Consider a periodic potential $V_-(x)$ whose ground state energy is zero, eigenvalues are $E_n^{(-)}$ and eigenfunctions (band edges) are $\psi_n^{(-)}(x)$. Let the ground state wave function be denoted by $\psi_g(x) \equiv \psi_0^{(-)}(x)$. One constructs the superpotential $W(x)=-\psi'_g(x)/\psi_g(x)$. The original potential and its supersymmetric partner potential are then given by $V_{\pm}(x) = W^2(x) \pm W'(x)$. The eigenvalues are the same for both potentials and their un-normalized eigenfunctions are related by
\be
\psi_0^{(+)}(x) \propto 1/\psi_0^{(-)}(x)~~,~~\psi_n^{(+)}(x) 
\propto \left[ \frac{d}{dx}+W(x) \right]
\psi_n^{(-)}(x)~~,~~ n \ge 1~.
\ee
This technique is immediately applicable to PT-invariant potentials $V^{PT}(x)$ of the type given in eq. (\ref{eq20}) from which the ground state energy has been subtracted. The only caution to keep in mind is that the ground state of $V^{PT}(x)$ corresponds to the highest eigenstate of $V(x)$ as indicated by eq. (\ref{eq5}).

Let us apply the above formalism to the Lam\'e potentials. First, for the special
case $a=1$, one has 
$$
V_{-}(x)=2m\sn^2(x,m)-m~~,~~ [V^{PT}]_-(x)=-2m\sn^2(ix+\beta,m)+m+1~~,~~\psi_g(x)=\sn(ix+\beta,m)~,$$
\be
W^{PT}(x)=-i\frac{\cn(ix+\beta,m)~\dn(ix+\beta,m)}{\sn(ix+\beta,m)},
~[V^{PT}]_+(x)=-2m\sn^2(ix+\beta+iK'(m),m)+m+1.
\ee
Here, the result of invoking supersymmetry is basically a translation of the independent variable in $[V^{PT}]_-(x)$. Such potentials are usually called self-isospectral potentials \cite{dm}. 

For higher $a$ values, the two supersymmetric partner potentials are 
quite different in shape from each other \cite{ks1}, even though they both have the same band edge eigenvalues. 
Let us now explicitly consider the case $a=3$.
Here, using the ground state
eigenfunction of the PT-invariant potential $V^{PT}(x)=-12m\sn^2(ix+\beta,m)-E_g$, where $E_g$ 
is given in eq. (\ref{eq13}), 
we find that the corresponding superpotential is
\be\label{eq24}
W^{PT}(x) = -i\frac{\cn(ix+\beta,m) \dn(ix+\beta,m)}{\sn(ix+\beta,m)}
+ 10im \frac{\cn(ix+\beta,m) \sn(ix+\beta,m) \dn(ix+\beta,m)}
{[2+2m-\delta_3-5m\sn^2(ix+\beta,m)]}\,,
\ee
so that the supersymmetric partner potential is
$[V^{PT}]_{+} (x) = [W^{PT}(x)]^2  +[W^{PT}(x)]' \,.$ 
In this way, one has discovered another PT-invariant complex
potential with a finite number of band gaps. It is plotted in Figure 2.
 
We can obtain yet other analytically solvable, complex, PT-invariant 
potentials by exchanging the orders of applying anti-isospectral 
transformations and supersymmetry. For example, we could first determine the 
supersymmetric partner of a solvable associated 
Lam\'e potential and then compute the corresponding PT-invariant
potential. The results for the $a=1$ Lam\'e potential are:
\be
V_+(x)=2m\sn^2(x-K(m),m)-m~,~ [V_+]^{PT}(x) = -2m\sn^2(ix+\beta-K(m),m)+m+1~,
\ee
where the potentials have been adjusted so as to have zero ground state energy. Again, for this $a=1$ example, one finds that 
$[V_+]^{PT}(x)$ is essentially the same as $[V^{PT}]_+(x)$ with a constant complex shift of the independent variable $x$.

The situation is much richer and more interesting for Lam\'e potentials with higher $a$ values. For instance, it is
shown in ref. \cite{ks1} that the supersymmetric partner
potential of the $a=3$ Lam\'e potential is given by
\be\label{eq27}
V_{+}(x)=-12m\sn^2(x,m)+\frac{2m^2\,\sn^2(x,m)\cn^2(x,m)}{\dn^2(x,m)}
\frac{[2m+\delta_1+11-15m\sn^2(x,m)]^2}
{[2m+\delta_1+1-5m\sn^2(x,m)]^2}\,,
\ee
and the corresponding PT-invariant complex potential $[V_{+}]^{PT}$
is simply
obtained from here by using the anti-isospectral transformation
$x \rightarrow ix+\beta$ and subtracting off the ground state energy
$E_g$ given by eq. (\ref{eq24}) from it. This potential $[V_{+}]^{PT}$ is plotted in Figure 3.
Clearly its band edge energy 
eigenvalues are simply related to those of $V_{+}(x)$ and hence to the $a=3$
Lam\'e potential by relation (\ref{eq5}). Hence the band edge energy
eigenvalues of $[V_{+}]^{PT}$ are 
identical to those of $[V^{PT}]_{+}$ and $[V^{PT}]_{-}$ even though the three potentials are
distinct. For the example under consideration, this is just the statement that the three 
different complex PT-invariant potentials $[V^{PT}]_{-}(x), [V^{PT}]_{+}(x), [V_{+}]^{PT}(x) $ 
plotted in Figures 1,2,3 all have the same band structure.

In Table 2 we have given the expression for the band edge eigenvalues
and eigenfunctions for the PT-invariant complex potential
$V^{PT}(x)=-6m\sn^2(ix+\beta,m)-2m{\cn^2 (ix+\beta,m)}/{\dn^2(ix+\beta,m)}\,$
suitably adjusted by subtracting its ground state energy
$E_g = -5-m-2\sqrt{4-3m}$ so that the
lowest band edge is at zero energy. 
Using the ground state wave function of this potential, 
the corresponding superpotential turns out to be
\bea\label{eq28}
&&W^{PT}=i\frac{\sn(ix+\beta,m) \dn(ix+\beta,m)}{\cn(ix+\beta,m)}
-im\frac{\cn(ix+\beta,m) \sn(ix+\beta,m)}{\dn(ix+\beta,m)} \nonumber
\\
&&~~~~~~~~~~~~~~~~~~~~~~~~~-6im\frac{\sn(ix+\beta,m) \dn(ix+\beta,m) \cn(ix+\beta,m)}
{3m\sn^2(ix+\beta,m)-2+\sqrt{4-3m}}\,.
\eea
Hence the corresponding supersymmetric partner
potential $[V^{PT}]_{+} (x) = [W^{PT}(x)]^2  +[W^{PT}(x)]'$ is easily calculated. 
On the other hand, yet another PT-invariant 
potential with the same band edges can be obtained by starting 
from the partner potential of the $(a=2, b=1)$ associated Lam\'e potential 
$V(x)=6m\sn^2(x,m)+2m{\cn^2 (x,m)}/{\dn^2(x,m)}$, and applying the
anti-isospectral transformation. We get  
\be\label{eq29}
[V_{+}]^{PT}(x)=-2m\sn^2(ix+\beta,m)
-6m\frac{\cn^2 (ix+\beta,m)}{\dn^2(ix+\beta,m)}+5+m+2\sqrt{4-3m}\,.
\ee
It is worth noting that again the two potentials $[V_{+}]^{PT}$ and
$[V^{PT}]_{+}$ are quite different even though 
they have the same band edge eigenvalues.
Further, while the initial associated Lam\'e potential
is self-isospectral, its PT-transform is not so.
In fact,  this seems to be true in general. In particular, whereas the
associated Lam\'e potentials with $b=a-1$ are isospectral, we find that
the corresponding PT-invariant periodic potentials are not self isospectral
except when $a=1$. 

Finally, let us comment that for the PT-invariant potential $i\sin^{2N+1}(x)$, 
Bender et al. \cite{bdm} found that the band edge eigenfunctions are always 
$2\pi$ periodic and unlike other lattice problems, the anti-periodic 
band edge eigenfunctions of period $4\pi$ were absent. They speculated 
whether this could perhaps be a unique signal of PT symmetry. However, in this 
letter, we have seen many examples where this is not true. In particular, 
Table I shows an example with both periodic as well as anti-periodic
band edges, showing that the absence of anti-periodic band edges is not 
a general property of PT-invariant periodic potentials.

Acknowledgements: One of (US) would like to thank the U.S. Department of Energy and the International Centre for Theoretical Physics, Trieste, Italy for partial support for this research.

\vspace{0.5in}
\noindent{\Large \bf Figure Captions}
\vskip .5 true cm

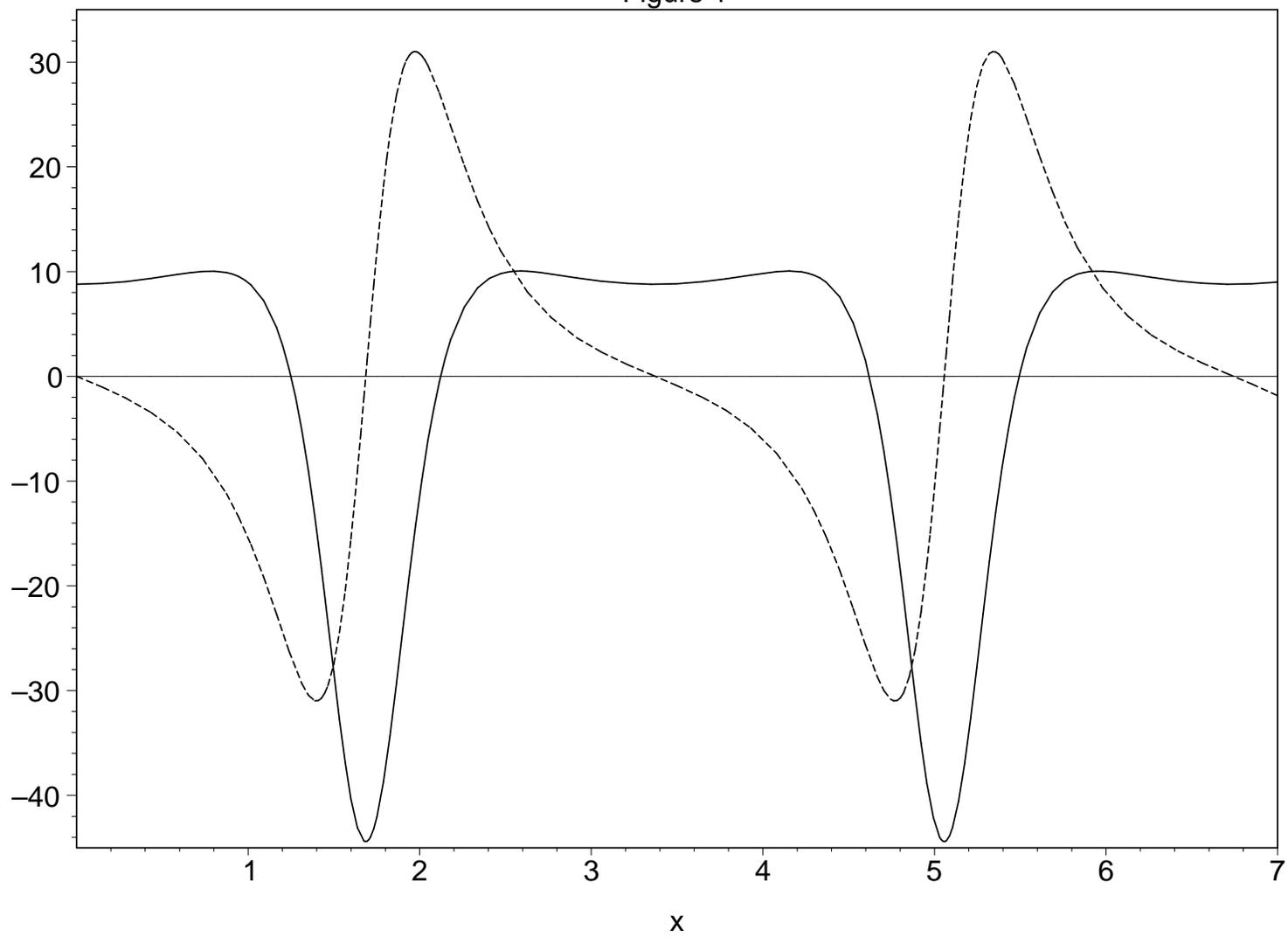
\noindent {\bf Figure 1:} A plot of the real and imaginary parts of the complex PT-invariant potential 
$[V^{PT}]_-(x) = -12 m \sn^2(ix+\beta, m)-E_g$, where $E_g=-5-5m-2\delta_3$, and $\delta_3 \equiv \sqrt{4 - 7m + 4m^2}$ (see ref. \cite{ks1}). The potential has been defined so as to have zero ground state energy. The plot is for the choice $m = 0.75,~ \beta = 0.5$. 
The potential has a period $2K'(0.75)=3.3715$. The continuous
curve denotes the real part and the dashed curve denotes the imaginary
part.\sss

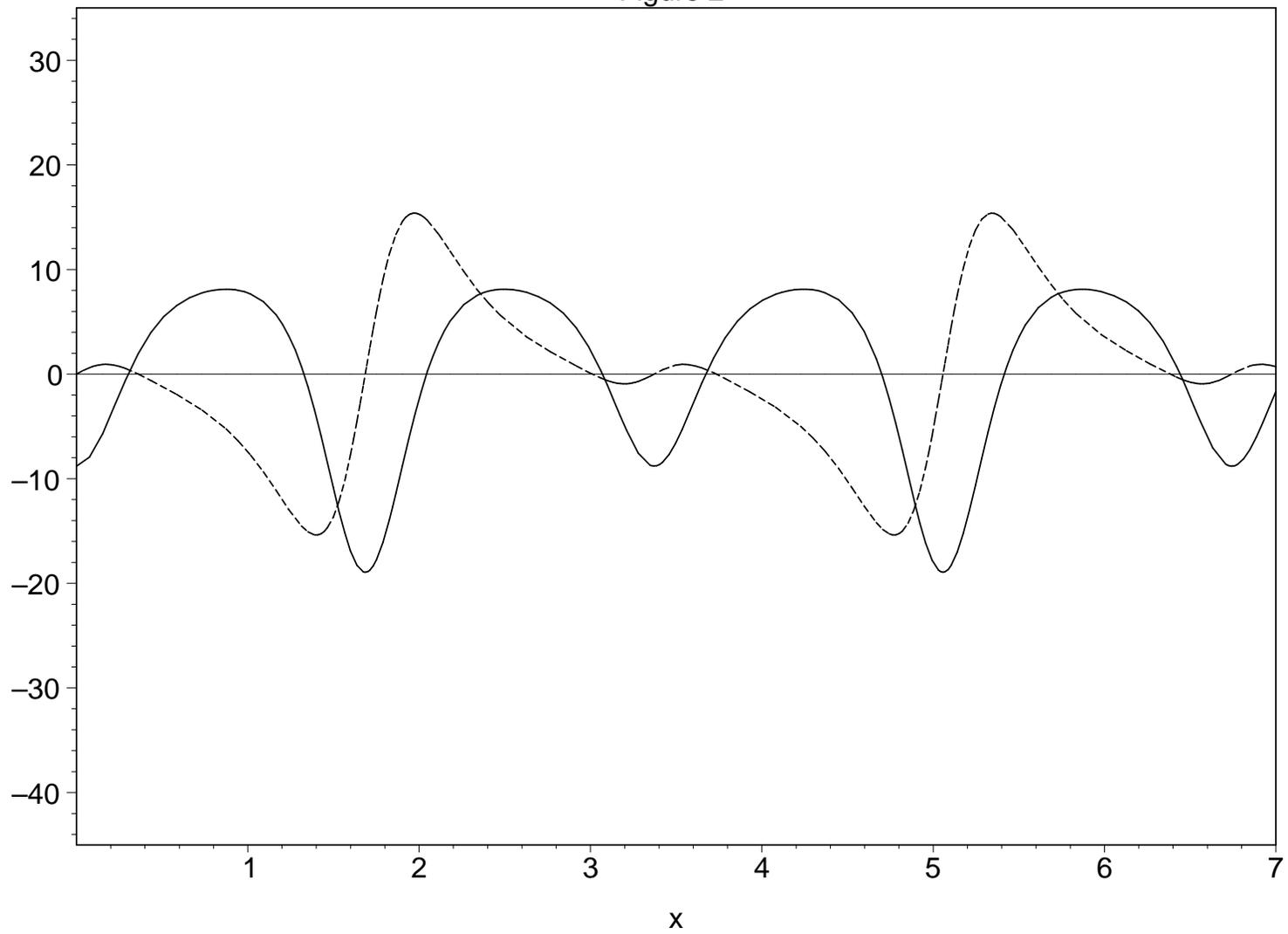
\noindent {\bf Figure 2:} A plot of the real (continuous curve) and 
imaginary (dashed curve) parts of the supersymmetric partner potential 
$[V^{PT}]_+(x)$ of the complex PT-invariant potential shown in Figure 1, 
for the choice $m = 0.75,~ \beta = 0.5$.\sss

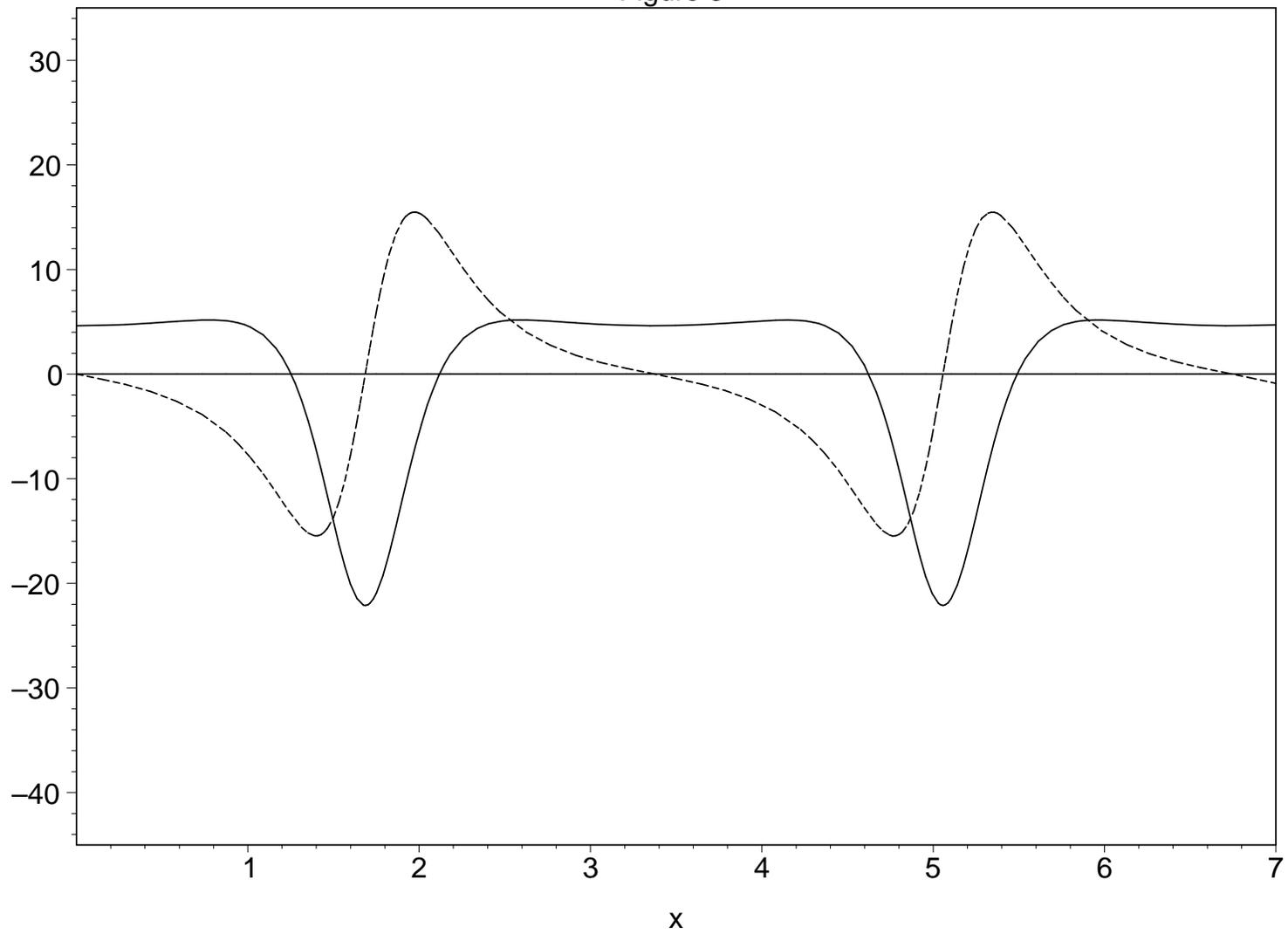
\noindent {\bf Figure 3:} A plot of the real (continuous curve) and 
imaginary (dashed curve) parts of the PT-invariant 
potential $[V_{+}]^{PT}(x)$ 
obtained by first taking the supersymmetric partner of the $a=3$ Lam\'e potential and then applying the anti-isospectral transformation $x \rightarrow x+i\beta$, for the choice $m = 0.75, ~\beta = 0.5$. The constant energy $-3-2\delta_3-2\delta_1$ (see Table 1) has been subtracted off, so that the ground state of $[V_{+}]^{PT}(x)$ has zero energy.

\newpage


\noindent{\bf Table 1:} The eigenvalues and eigenfunctions for the 7 band edges of the 
PT-invariant Lam\'e potential $[V^{PT}]_{-}(x) =-12m\sn^2(ix+\beta,m)-E_g$,
where $E_g=-5-5m-2\delta_3$ [eq. (\ref{eq13})], and $\delta_1 \equiv \sqrt{1-m+4m^2}, 
~\delta_2 \equiv \sqrt{4-m+m^2}, ~\delta_3 \equiv \sqrt{4-7m+4m^2}$. 
The potential has a period $2K'(m)$. 
The real periods of various eigenfunctions are also tabulated. \sss

\oddsidemargin      -0.2in

\bigskip
\begin{tabular}{cccccc}
\hline
 $E$ & $\psi^{(-)}$ &  
${\rm Period}$\\
\hline
 $0$ & $\sn (ix+\beta,m) [2+2m -\delta_3 -5m {\rm sn}^2 (ix+\beta,m)]$
& $2K'(m)$\\
 $ 3m + 2\delta_3-2\delta_2$ & ${\cn}(ix+\beta,m)[2+m-\delta_2 -5m
\sn^2 (ix+\beta,m)]$
& $4K'(m)$ \\
 $ 3+ 2\delta_3-2\delta_1$ & ${\rm dn}(ix+\beta,m)[1+2m-\delta_1 -5m
\sn^2 (ix+\beta,m)]$
& $4K'(m)$ \\
 $ 1+m+2\delta_{3}$ & ${\rm sn}(ix+\beta,m) \cn (ix+\beta,m) {\rm
dn}(ix+\beta,m)$
& $2K'(m)$ \\
 4$\delta_3$ & $\sn (ix+\beta,m) [2+2m +\delta_3 -5m {\rm sn}^2 (ix+\beta)]$ 
& $2K'(m)$ \\
 $ 3m + 2\delta_3+2\delta_2$ & ${\cn}(ix+\beta,m)[2+m+\delta_2 -5m
\sn^2 (ix+\beta,m)]$
& $4K'(m)$ \\
 $ 3+ 2\delta_3+2\delta_1$ & ${\rm dn}(ix+\beta,m)[1+2m+\delta_1 -5m \sn^2
(ix+\beta,m)]$
& $4K'(m)$ \\
\hline
\end{tabular}
\bigskip

\vskip 1.8 true cm
%

\noindent {\bf Table 2:} The eigenvalues and eigenfunctions for the 5 band edges of
the PT-invariant associated Lam\'e potential $[V^{PT}]_{-}(x) = -6m\sn^2(ix+\beta,m)
-2m{\cn^2(ix+\beta,m)}/{\dn^2(ix+\beta,m)}-E_g$, where $E_g$ is 
given by eq. (\ref{eq21}). Here $\delta_4 \equiv \sqrt{4-5m+m^2}$. 
The potential has a period $2K'(m).$ The real periods of various
eigenfunctions are also tabulated. \sss

%
\bigskip
\begin{tabular}{cccccc}
\hline
 $E$ & $\psi^{(-)}$ & 
${\rm Period}$\\
\hline
 $0$ & $\frac{\cn (ix+\beta,m)}{\dn
(ix+\beta,m)}[3m\sn^2(ix+\beta,m)-2+\sqrt{4-3m}]$  
 & $2K'(m)$\\
 $ 2\sqrt{4-3m}-m-2\delta_4 $ & ${\sn (ix+\beta,m)\over \dn
(ix+\beta,m)} [3m{\sn}^2 (ix+\beta,m) -2-m+\delta_4]$
 & $4K'(m)$\\
 $ 2\sqrt{4-3m}-m+2\delta_4 $ & ${\sn (ix+\beta,m)\over \dn
(ix+\beta,m)} [3m{\sn}^2 (ix+\beta,m) -2-m-\delta_4]$
 & $4K'(m)$\\
 $ 4\sqrt{4-3m} $ & ${\cn (ix+\beta,m)\over \dn
(ix+\beta,m)} [3m{\sn}^2 (ix+\beta,m) -2-\sqrt{4-3m}]$
 & $2K'(m)$\\
 $ 5-3m+2\sqrt{4-3m}$ & ${\rm dn}^2 (ix+\beta,m)$
 & $2K'(m)$\\
\hline
\end{tabular}
\bigskip

\vskip 2.2 true cm
\end{document}